\documentclass[sigconf,nonacm]{acmart}

\usepackage[capitalise]{cleveref}
\usepackage{amsmath}

\AtBeginDocument{%
  }

\copyrightyear{2024}
\acmYear{2024}
\setcopyright{rightsretained}
\acmConference[RecSys '24]{18th ACM Conference on Recommender Systems}{October 14--18, 2024}{Bari, Italy}
\acmBooktitle{18th ACM Conference on Recommender Systems (RecSys '24), October 14--18, 2024, Bari, Italy}
\acmDOI{10.1145/3640457.3688163}
\acmISBN{979-8-4007-0505-2/24/10}

\newcommand{\IALS}{\textit{IALS}}
\newcommand{\MFNS}{\textit{BPR}}
\newcommand{\moved}{\textit{MoveUp}}
\newcommand{\fair}{\textit{FAIR}}
\newcommand{\lambdafive}{$\lambda5$}
\newcommand{\lambdaseven}{$\lambda7$}

\newcommand{\DeterministicTopN}{\textit{Det}}
\newcommand{\RandomTopN}{\textit{Rnd}}
\newcommand{\InspectionAbandon}{\textit{IA}}
\newcommand{\InspectionAbandonBiasedProMale}{\textit{Biased}}

\begin{document}

\title{It's Not You, It's Me: The Impact of Choice Models and Ranking Strategies on Gender Imbalance in Music Recommendation}

\author{Andres Ferraro} 
\email{andresferraro@acm.org} 
\orcid{0000-0003-1236-2503}
\affiliation{%
  \institution{Pandora-Sirius XM}
  \city{Oakland}
  \state{CA}
  \country{USA}
}

\author{Michael D. Ekstrand}
\email{mdekstrand@drexel.edu} 
\orcid{0000-0003-2467-0108}
\affiliation{
\institution{Drexel University}
\city{Philadelphia}
\state{PA}
\country{USA}
}

\author{Christine Bauer}
\email{christine.bauer@plus.ac.at}
\orcid{0000-0001-5724-1137}
\affiliation{%
  \institution{Paris Lodron University Salzburg}
  \city{Salzburg}
  \country{Austria}
}

\begin{abstract}
      As recommender systems are prone to various biases, mitigation approaches are needed to ensure that recommendations are fair to various stakeholders.
      One particular concern in music recommendation is artist gender fairness.
      Recent work has shown that the gender imbalance in the sector translates to the output of music recommender systems, creating a feedback loop that can reinforce gender biases over time. 

      In this work, we examine that feedback loop to study whether algorithmic strategies or user behavior are a greater contributor to ongoing improvement (or loss) in fairness as models are repeatedly re-trained on new user feedback data.
      We simulate user interaction and re-training to investigate the effects of ranking strategies and user choice models on gender fairness metrics.
      We find re-ranking strategies have a greater effect than user choice models on recommendation fairness over time.
\end{abstract}

\keywords{User choice models, re-ranking, artists, music, 
gender, fairness, bias} 
\maketitle

\section{Introduction}\label{sec:introduction}
Recommender systems significantly impact users' activities on a wide range of platforms.
Streaming platforms have become one of the primary sources of music consumption~\citep{IFPI2024}. 
Typically, such platforms integrate music recommender systems (MRS) that learn from large-scale user behavior and music features~\citep{schedl2022_handbook} to recommend music (at various levels, including songs, artists, etc.) tailored to a specific user.
With the prevalence and impact of automatic recommendation in music (and other domains), it is vital to consider the \textit{fairness} of both the recommendations themselves and their interaction with users and the broader sociotechnical context~\citep{Ekstrand2022Fairness}.

In this work, we specifically examine biases related to artist gender. Gender imbalance is a highly topical subject in the music sector~\citep[e.g.][]{smith2022_inclusion_recording,smith2024_inclusion_recording,wang2019gender,watsonprogramming}. 
From interviews with artists, \citet{Ferraro2021Fair} learn that artists care about the gender imbalance in the music industry, a finding partially supported by interviews in \citet{Dinnissen2023Amplifying} as well.
As artists consider MRS a potential solution to promote content by female artists to reach a gender balance in what users consume~\citep{Ferraro2021Fair}, \citet{ferraro2021_break} analyze MRS approaches regarding gender bias and propose bias mitigation strategies to counteract the gender imbalance. In a simulation study, they demonstrate that gradually increasing exposure of underrepresented gender group can interrupt long-term bias amplification.
In a field study on Spotify, \citet{EppsDarling2020Gender} found that increasing the fraction of tracks by female artists in algorithmic music recommendations resulted in continuing increase in the prevalence of tracks by female artists in users' ``organic'' listening (selecting and playing songs without the recommender).
Compared to other domains, music has a particularly biased starting point with respect to gender in the underlying data because there is a substantial gender gap: male artists dominate the field, while female and nonbinary artists form a smaller group, though some within this minority group are highly popular~\citep{smith2022_inclusion_recording,smith2024_inclusion_recording}.

In this work, we build on the findings of \citet{ferraro2021_break} and \citet{bauer2023_strategies_murs}, using a similar simulation approach to explore the effects of different post-processing strategies for bias mitigation.
In particular, we expand those results to additional post-processing strategies and examine the effects of two different elements of the feedback loop: the \textbf{re-ranking strategy} to augment the fairness of the recommendations, and the \textbf{user choice model} that encodes how users select results from recommendations. This allows us to examine whether algorithmic strategies or user behavior are a greater contributor to ongoing improvement (or loss) in fairness.
We note that this study is purely descriptive---we make no claims about what the balance of artist genders in recommendation should be, but seek to document how that balance is affected by various elements of the sociotechnical feedback loop.
While our work focuses specifically on gender imbalance in the music industry, our contribution is relevant to other provider characteristics and content domains, such as books~\citep{Ekstrand2021}.

\section{Related Work}\label{sec:related}

There is a wealth of research on biases in recommender systems. 
Popularity bias has been widely researched for many years~\citep[e.g.][]{elahi2021investigating,Kowald2022_Popularity, ekstrandSturgeonCoolKids2017, celmaHitsNichesHow2008}. More recently, research on bias and bias mitigation have been particularly addressed from the perspective of societal fairness: are users or item providers treated fairly---and if not, how can we create fair(er) recommender systems?
Biases in recommender system behavior and outputs can arise from many different sources, including the algorithms themselves, corpus data, training data, and users' ongoing interactions with the system~\citep{Ekstrand2022Fairness}.
Unfairness arising from both underlying data and recommendation models, and their interplay, has been documented in prior studies~\citep[e.g.][]{ferraro2021_break,Ekstrand2021}.

In this paper, we are concerned with \textit{fair exposure} or \textit{visibility} for artists across genders. 
Re-ranking methods are widely adopted to improve such outcome fairness~\citep{Wang2023_fairness, sonboliOpportunisticMultiaspectFairness2020a, rajOptimizingRankingGridlayout2024, EppsDarling2020Gender,ferraro2021_break,Sonboli2020opportunistic}, post-processing the outputs of standard recommendation models to improve the fairness of the resulting lists (or, in some cases, grids or other displays). 
In the context of gender (im)balance, these strategies can be deployed both to improve equity in the recommendation themselves---increasing the exposure that items created by female and 
gender-minority artists receive---and also to influence future consumption patterns towards gender parity~\citep{EppsDarling2020Gender}.

Most work on fair recommendation is concerned with fairness at a single point in time~\citep[e.g.][]{Ekstrand2021}; \citet{EppsDarling2020Gender} looked at the recommender impact on users but did not close the loop for user impact on subsequent recommendations. Real recommender systems, however, are iterative systems with a feedback loop between the system and its users.
Simulation is a useful mechanism for studying feedback loops: \citet{Mansoury2020Feedback} and \citet{jannach2015recommenders} 
use simulations to study popularity bias effects, \citet{zhang2020consumption} find that users' high reliance on recommender systems provides suboptimal performance outcomes in the long run, and \citet{chaney2018algorithmic} document longitudinal convergence in recommendations between users.
While relying on the same simulation approach, \citet{ferraro2019} study exposure biases across artists of different styles, whereas \citet{ferraro2021_break} and \citet{bauer2023_strategies_murs} address gender imbalances in artist exposure. .

Many simulations use very simple models of user response, e.g., assuming that a user would consume the top~$k$ items~\citep{ferraro2021_break}.
Recent research on user choice models~\citep{Hazrati2024_choicemodels} shows that different choice models can influence the overall choice distribution and performance of the recommenders. Further, different recommender algorithms affect users' choices in different ways~\citep{HAZRATI2022102766}.  
It is therefore necessary to employ richer models of user action to study recommender system feedback loops.

Few works compare the relative effects of recommendation algorithms and user choice models. One example is that of
\citet{Fabbri_Croci_Bonchi_Castillo_2022}, who studied exposure inequality in people recommender systems. Their stochastic user choice models did not impact the exposure as much as the recommender algorithm and the underlying social graph structure.

Understanding the relative impact of recommendation techniques and user responses remains a gap in understanding the fairness of recommender systems and developing robust, empirically-grounded tools for ensuring recommendation contributes to a fair and equitable society. 
In this paper, we build on prior work by two of the authors~\citep{ferraro2021_break, bauer2023_strategies_murs} that used simulation to study the imbalance in gender exposure in MRS; we extend their methods to study and compare the impact of base recommendation models, re-ranking strategies, and user choice models on gender balance in an iterative setting where recommendation models are repeatedly re-trained in response to user feedback.

\section{Methods}\label{sec:methods}
\begin{figure*}
\centering
\includegraphics[width=\textwidth]{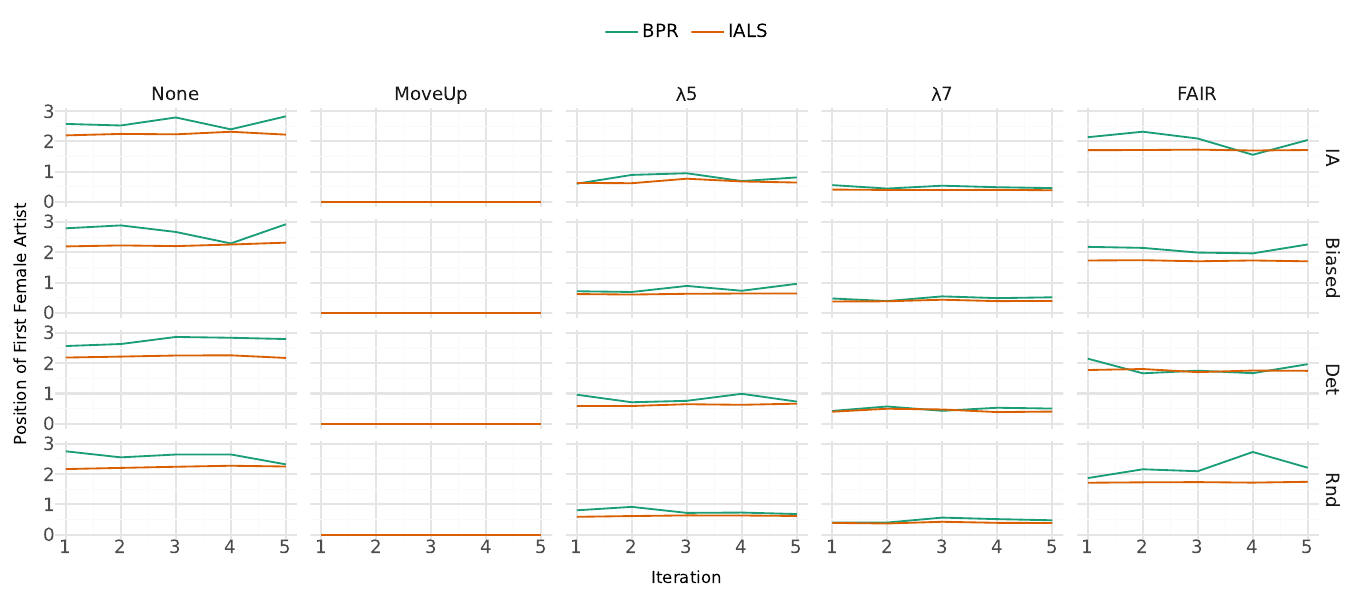}
\caption{Average position of the first female artist ($PFA$) over iterations. Position~$0$ is the highest rank.}
\label{fig:first_female}
\end{figure*}

\begin{figure*}
\centering
\includegraphics[width=\textwidth]{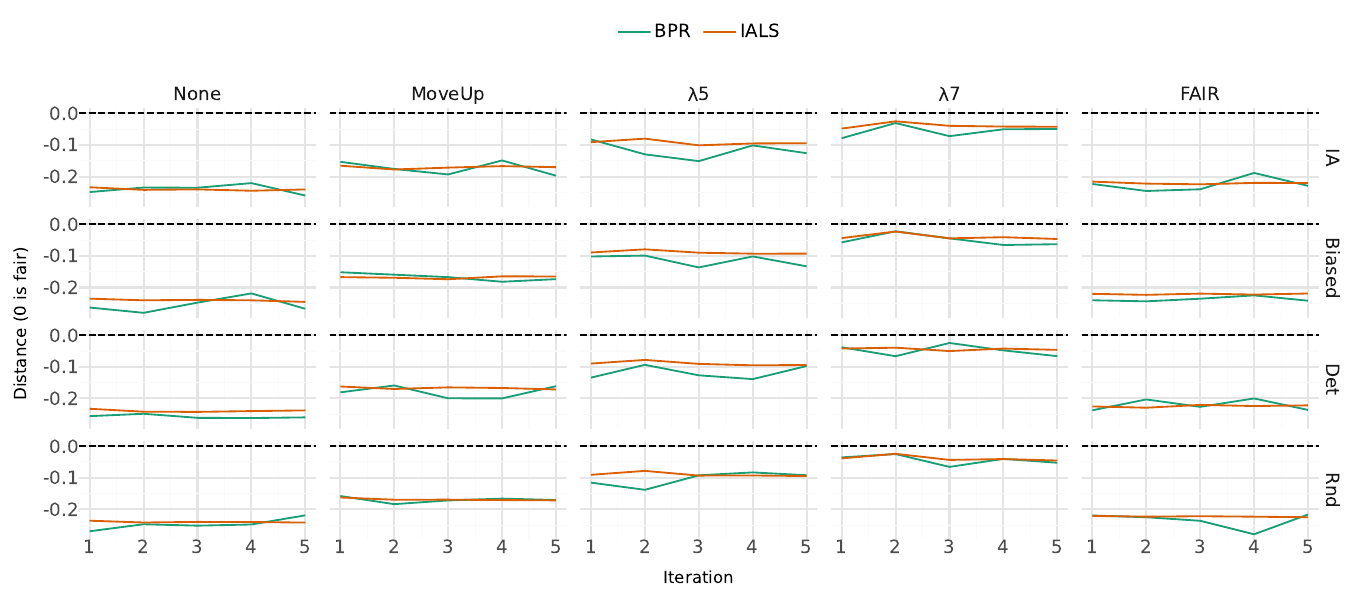}
\caption{$AWRF$ over iterations. $0$~is fair; negative values mean that unprotected-group (male) artists are overexposed.}
\label{fig:awrf}
\end{figure*}

\begin{figure*}
\centering
\includegraphics[width=\textwidth]{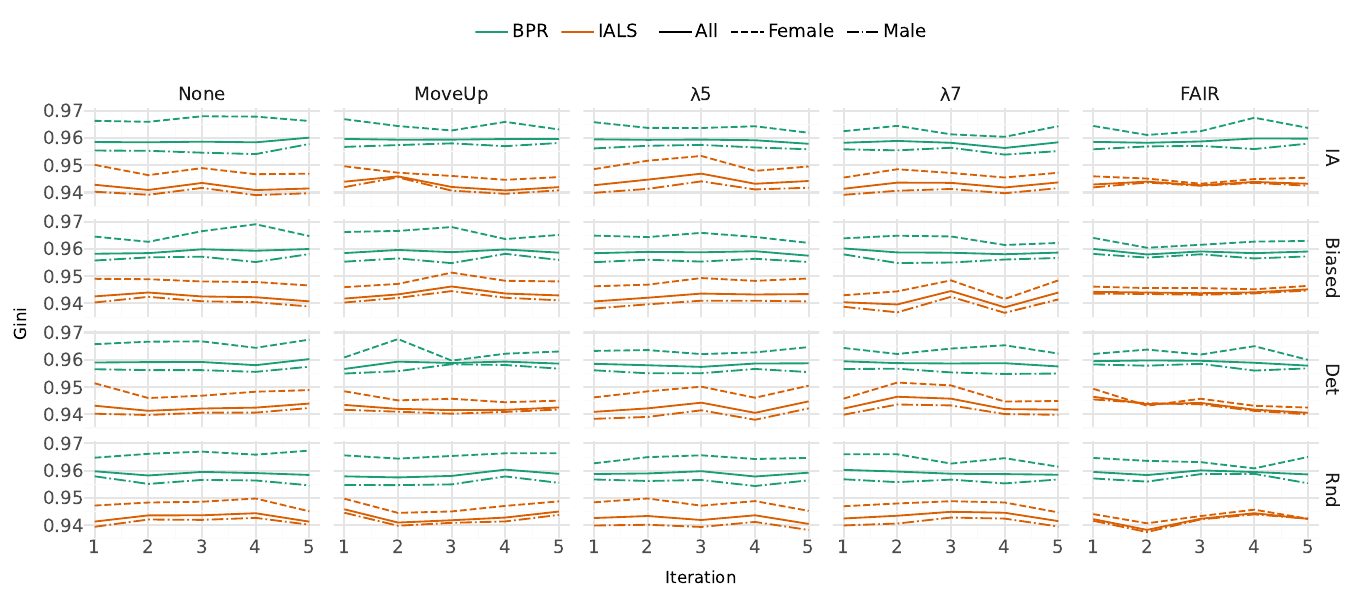}
\caption{Gini coefficients ($Gini@10$) over iterations, both overall and for male and female artists. $1$~is maximum inequality.}
\label{fig:gini}
\end{figure*}

We build our experiment on a subset of the \textit{LFM-2b} dataset~\citep{schedl2022_lfm2b}.
We use two base recommendation models,
\IALS{} (implicit-feedback matrix factorization with alternating least squares~\citep{Hu2008_ials})
and \MFNS{} (matrix factorization trained with pairwise rank loss~\citep{rendleBPRBayesianPersonalized2009}),
to generate initial recommendation lists and reprocess these lists through various post-processing bias mitigation strategies. 
We ran the experiments with LensKit~\citep{ekstrand2020_lenskit}, using LensKit's \IALS{} implementation and a PyTorch \MFNS{} implementation for LensKit.
We tuned the hyperparameters for both base models using random search to optimize $MRR$ over $1000$-item rankings on a test set.

We first describe the mitigation strategies we study, then remaining details of the data, metrics, and experimental simulation.\footnote{Code available at \url{https://doi.org/10.5281/zenodo.13315571}.}

\subsection{Approaches}

We choose \IALS{} and \MFNS{} as the basis approaches for our analysis because these are well-known algorithms for collaborative filtering. Each mitigation strategy post-processes the ranking produced by \IALS{} or \MFNS{} to improve the gender fairness of the final ranking.

We test three strategies for recommendation post-processing:
\begin{enumerate}
\item[]\textbf{\moved{}} Move the first item by a female artist to the first rank.
\item[]\textbf{$\boldsymbol{\lambda{}5, \lambda{}7}$}  Penalize items by male artists by moving each of them $\lambda$ positions downward in the ranking~\citep{ferraro2021_break}; based on the original paper, we use $\lambda=5$ and $\lambda=7$. 
\item[]\textbf{\fair{}} The FA*IR ranking algorithm from \citet{zehlikeFAIRFair2017}: For each position, select the highest-scored item in the original ranking that will not cause the protected group (non-male artists) to be statistically significantly underrepresented concerning the target proportion (which we set to $50\%$).
\end{enumerate} 
\begin{description}
\item[\moved{}] Move the first item by a female artist to the first rank.
\item[$\boldsymbol{\lambda{}5, \lambda{}7}$]  Penalize items by male artists by moving each of them $\lambda$ positions downward in the ranking~\citep{ferraro2021_break}; based on the original paper, we use $\lambda=5$ and $\lambda=7$. 
\item[\fair{}] The FA*IR ranking algorithm from \citet{zehlikeFAIRFair2017}: For each position, select the highest-scored item in the original ranking that will not cause the protected group (non-male artists) to be statistically significantly underrepresented concerning the target proportion (which we set to $50\%$).
\end{description} 

\IALS{} and \MFNS{} without any adaptations (\textit{None}) serve as baselines. 

\subsection{Dataset}
For our exploratory analysis, we rely on a subset of interactions between 2013--2020 in the \textit{LFM-2b} dataset~\citep{schedl2022_lfm2b} enriched with artist gender information collected from MusicBrainz.org (\textit{MB})\footnote{\url{https://musicbrainz.org}}. 
We limit the data to only `solo' artists---where the artist is a person---for which \textit{MB} reports the gender identities. We consider the categories female, male, and nonbinary\footnote{Nonbinary refers to a broad spectrum of different individual identities. From \textit{MB}, we merged the categories nonbinary and other.}.

The final data contains $436,789$ artists, of which $93,316$ are categorized as female, $342,523$ as male, and $950$ as nonbinary.
Following the procedure of \citet{ferraro2021_break}, we use a 15-core of user-artist interactions, retaining $78,021$~users and $187,471$~artists, of whom $21.675\%$ represent female artists.
For hyperparameter tuning, we split data by time to get $90\%$ training and $10\%$ test data, discarding test users with no training ratings.

\subsection{Metrics}\label{sec:metrics}

We use several metrics to understand the system's behavior from different perspectives:
\begin{itemize}

    \item \textit{First-position exposure}.
We focus on the position in the recommendation rankings because users interact more frequently with only the top-ranked items (i.e., position bias)~\citep{collins2018position}. To this end---in line with \citet{ferraro2021_break}, we average for each user the \textit{position of the first female artist} ($PFA$) in the recommendation ranking, with the first position as~$0$.

    \item \textit{Overall exposure} with $AWRF$ (Attention-Weighted Rank Fairness \citep{AWRF, raj2022:ranking}). This uses rank-discounting (as in metrics like $nDCG$ and $RBP$; we followed TREC in using $nDCG$'s logarithmic discounting~\citep{TREC2021}) to estimate the exposure value of each rank position, and measuring the fairness of exposure provided to each group.
    As with \fair{}, we treat male artists as unprotected, and measure fairness by the difference between the exposure distribution in the ranking and a target of $50\%$, and average over all recommendation lists.

\item \textit{Diversity}.
We use the \textit{Gini index} ($Gini@k$) to measure how concentrated the recommendations are on a few artists both overall and disaggregated by gender.
$Gini@k=1$ indicates that all recommendations go to the same artist, while $0$ means all artists are equally recommended.
\end{itemize}

\subsection{User Choice Models}
In the simulation, we consider four different user choice models (each implemented for $N=10$):
\begin{enumerate}
\item[]\textbf{Deterministic (\DeterministicTopN{})}
    User listens to all~$N$ recommended items. 
\item[]\textbf{Random (\RandomTopN{})}
    For each item in a user's top~$N$ recommended items, the user listens with probability~$0.5$.
\item[]\textbf{InspectionAbandon (\InspectionAbandon{})} A probabilistic cascade model~\citep{rajUnifiedBrowsingModels2023}: for each item in the top-$N$ recommendations, with probability~$0.5$ the user listens to that item. After
    listening or ignoring the item, they stop listening entirely with probability~$0.3$; otherwise, they continue to the next item.
    The \InspectionAbandon{} model is based on cascade models from the existing evaluation  literature~\citep{rajUnifiedBrowsingModels2023}, implementing that users have a higher probability of consuming higher-ranked items than lower-ranked ones.
\item[]\textbf{\InspectionAbandonBiasedProMale{}} Variant of \InspectionAbandon{} in which users are biased so that they are $10\%$~more likely to listen to male artists than others.
\end{enumerate}
\begin{description}
\item[Deterministic (\DeterministicTopN{})]
    User listens to all~$N$ recommended items. 
\item[Random (\RandomTopN{})]
    For each item in a user's top~$N$ recommended items, the user listens with probability~$0.5$.
\item[InspectionAbandon (\InspectionAbandon{})] A probabilistic cascade model~\citep{rajUnifiedBrowsingModels2023}: for each item in the top-$N$ recommendations, with probability~$0.5$ the user listens to that item. After
    listening or ignoring the item, they stop listening entirely with probability~$0.3$; otherwise, they continue to the next item.
    The \InspectionAbandon{} model is based on cascade models from the existing evaluation  literature~\citep{rajUnifiedBrowsingModels2023}, implementing that users have a higher probability of consuming higher-ranked items than lower-ranked ones.
\item[\InspectionAbandonBiasedProMale{}] Variant of \InspectionAbandon{} in which users are biased so that they are $10\%$~more likely to listen to male artists than others.
\end{description} 

\subsection{Simulation}
We simulate a user-recommender feedback loop to study the long-term effect of these mitigation strategies and user choice models. 
For each user, we use the choice model to select artists from the system's recommendations and record additional listening events for the selected artists.
We then retrain the model on this augmented data and compute recommendations for the next iteration.
We repeat this procedure for $5$~iterations, as the effects anchored after $5$~iterations in previous works~\citep{ferraro2021_break}.

\section{Results}\label{sec:results}

In this section, we present the results concerning the \textit{position of the first female artist} ($PFA$), $AWRF$, and the \textit{Gini coefficients}. We compare the re-ranking approaches (in the columns of the facet plots in the figures) and the user choice models (in the rows of the facet plots). 
While it is important to include the full gender spectrum in the data, model, and simulation, our data does not contain sufficient nonbinary artists to draw scientific conclusions from their inclusion in the analysis, so we report results only on \textit{male} and \textit{female} artists.

\cref{fig:first_female} clearly indicates that the re-ranking strategy determines the position of the first female artist, whereas the user choice model does not impact the ranking position of the first female artist. \IALS{} is more stable than \MFNS{}, showing almost no variation across iterations or choice models.
As \moved{} always puts a female artist on the very first position, it optimizes this metric by design. \lambdaseven{}---followed by \lambdafive{}---ranks the first female artist only on a slightly lower rank. \fair{} ranks the first female artist only slightly higher than the baselines without re-ranking (\textit{None}).

$AWRF$ is mostly consistent with $PFA$.
\cref{fig:awrf} indicates that the re-ranking strategies have more impact than user choice models on $AWRF$ (there is more variation between columns than rows), with little change over time.
Further, we observe that \fair{} contributes least to improving $AWRF$ compared to the baselines; moving only the first female artist (\moved{}) is more effective at overall exposure fairness even though it only adjusts the position of a single item. The $\lambda$-re-rankers were the most effective, with \lambdaseven{} improving exposure fairness the most.
As with $PFA$, \IALS{}-based recommendations had more stable gender exposure balance.

\cref{fig:gini} shows diversity as measured by $Gini@10$.
\IALS{} is more equitable than \MFNS{} across all re-ranking strategies and choice models.
Further, comparing $Gini@10_{female}$ and $Gini@10_{male}$ shows more equitable results for male than female artists: not only are female artists under-recommended (see $AWRF$), the exposure that does go to female artists is more concentrated on a smaller fraction of those artists.
With \fair{}, the gender gap in inequity closes, particularly with the \IALS{} base model.

\section{Discussion and Conclusion}\label{sec:discussion}

Our results show that re-ranking strategies have a greater effect than user choice models on the provider fairness of recommendations as the model is retrained with new user interactions over time: when it comes to longitudinal fairness, `it's not you' (the users), `it's me' (the recommender system). 
This effect is consistent across multiple metrics and underlying recommendation models.
That is, the model of user choice and response to recommendation had little effect on the fairness metrics we considered. 
Therefore, re-ranking strategies may be a useful tool for intervening in a biased world and addressing `societal imbalance'~\citep{mitchellAlgorithmicFairnessChoices2021}, as the algorithms have a stronger impact than the users' behavior in choosing items.
The base recommendation models we considered did exhibit different behavior, with \IALS{} delivering more stable results while \MFNS{} showed greater variation in fairness and diversity metrics, although without clear trends.
While the re-ranking strategies can break the original feedback loop, \IALS{} seems to anchor a `new' feedback loop within the $5$~iterations observed in our study.

It is not clear to what extent \MFNS{}'s volatility is a response to user data vs. noise and sensitivity to randomness in training.
Isolating these effects would require repeated re-runs of the simulation, which is computationally expensive; developing sample-efficient ways of identifying sources of noise~\citep{ekstrand2024distributions} is an important future research direction for recommender system simulation.

There are several important directions to extend and improve this work.
For one, while we considered nonbinary gender identities, the small number of nonbinary artists makes meaningful analysis with current methods difficult.
We plan future work to explore additional re-ranking strategies and choice models (including more heavily biased models), additional data sets and domains, and a broader range of recommendation strategies (including content-based and hybrid models). 
It is also vital in the future to examine intersectional impacts on artists who belong to multiple marginalized groups.

\begin{acks}
This publication was supported by the Excellence in Digital Sciences and Interdisciplinary Technologies (EXDIGIT) project, funded by Land Salzburg under grant number 20204-WISS/263/6-6022, and by the U.S. National Science Foundation under grant number 24-15042.
\end{acks}

\bibliographystyle{ACM-Reference-Format}
\bibliography{artists}


\begin{thebibliography}{41}


\ifx \showCODEN    \undefined \def \showCODEN     #1{\unskip}     \fi
\ifx \showDOI      \undefined \def \showDOI       #1{#1}\fi
\ifx \showISBNx    \undefined \def \showISBNx     #1{\unskip}     \fi
\ifx \showISBNxiii \undefined \def \showISBNxiii  #1{\unskip}     \fi
\ifx \showISSN     \undefined \def \showISSN      #1{\unskip}     \fi
\ifx \showLCCN     \undefined \def \showLCCN      #1{\unskip}     \fi
\ifx \shownote     \undefined \def \shownote      #1{#1}          \fi
\ifx \showarticletitle \undefined \def \showarticletitle #1{#1}   \fi
\ifx \showURL      \undefined \def \showURL       {\relax}        \fi
\providecommand\bibfield[2]{#2}
\providecommand\bibinfo[2]{#2}
\providecommand\natexlab[1]{#1}
\providecommand\showeprint[2][]{arXiv:#2}

\bibitem[Bauer and Ferraro(2023)]%
        {bauer2023_strategies_murs}
\bibfield{author}{\bibinfo{person}{Christine Bauer} {and}
  \bibinfo{person}{Andres Ferraro}.} \bibinfo{year}{2023}\natexlab{}.
\newblock \showarticletitle{Strategies for Mitigating Artist Gender Bias in
  Music Recommendation: a Simulation Study}. In
  \bibinfo{booktitle}{\emph{Proceedings of the Music Recommender Systems
  Workshop}} (Singapore) \emph{(\bibinfo{series}{MuRS 2023})}.
  \bibinfo{publisher}{Zenodo}, \bibinfo{numpages}{5}~pages.
\newblock
\urldef\tempurl%
\url{https://doi.org/10.5281/zenodo.8372477}
\showDOI{\tempurl}


\bibitem[Celma and Cano(2008)]%
        {celmaHitsNichesHow2008}
\bibfield{author}{\bibinfo{person}{{\`O}scar Celma} {and}
  \bibinfo{person}{Pedro Cano}.} \bibinfo{year}{2008}\natexlab{}.
\newblock \showarticletitle{From Hits to Niches? Or How Popular Artists Can
  Bias Music Recommendation and Discovery}. In
  \bibinfo{booktitle}{\emph{Proceedings of the 2nd {KDD} Workshop on
  Large-Scale Recommender Systems and the Netflix Prize Competition}} (Las
  Vegas, NV, USA) \emph{(\bibinfo{series}{NETFLIX '08})}.
  \bibinfo{publisher}{Association for Computing Machinery},
  \bibinfo{address}{New York, NY, USA}, Article \bibinfo{articleno}{5},
  \bibinfo{numpages}{8}~pages.
\newblock
\urldef\tempurl%
\url{https://doi.org/10.1145/1722149.1722154}
\showDOI{\tempurl}


\bibitem[Chaney et~al\mbox{.}(2018)]%
        {chaney2018algorithmic}
\bibfield{author}{\bibinfo{person}{Allison J.~B. Chaney},
  \bibinfo{person}{Brandon~M. Stewart}, {and} \bibinfo{person}{Barbara~E.
  Engelhardt}.} \bibinfo{year}{2018}\natexlab{}.
\newblock \showarticletitle{How Algorithmic Confounding in Recommendation
  Systems Increases Homogeneity and Decreases Utility}. In
  \bibinfo{booktitle}{\emph{Proceedings of the 12th {ACM} Conference on
  Recommender Systems}} (Vancouver, BC, Canada) \emph{(\bibinfo{series}{RecSys
  '18})}. \bibinfo{publisher}{Association for Computing Machinery},
  \bibinfo{address}{New York, NY, USA}, \bibinfo{pages}{224--232}.
\newblock
\urldef\tempurl%
\url{https://doi.org/10.1145/3240323.3240370}
\showDOI{\tempurl}


\bibitem[Collins et~al\mbox{.}(2018)]%
        {collins2018position}
\bibfield{author}{\bibinfo{person}{Andrew Collins}, \bibinfo{person}{Dominika
  Tkaczyk}, \bibinfo{person}{Akiko Aizawa}, {and} \bibinfo{person}{Joeran
  Beel}.} \bibinfo{year}{2018}\natexlab{}.
\newblock \showarticletitle{Position Bias in Recommender Systems for Digital
  Libraries}. In \bibinfo{booktitle}{\emph{Transforming Digital Worlds}}
  (Sheffield, UK) \emph{(\bibinfo{series}{iConference 2018})}.
  \bibinfo{publisher}{Springer}, \bibinfo{address}{Cham, Switzerland},
  \bibinfo{pages}{335--344}.
\newblock
\showISBNx{978-3-319-78105-1}
\urldef\tempurl%
\url{https://doi.org/10.1007/978-3-319-78105-1_37}
\showDOI{\tempurl}


\bibitem[Dinnissen and Bauer(2023)]%
        {Dinnissen2023Amplifying}
\bibfield{author}{\bibinfo{person}{Karlijn Dinnissen} {and}
  \bibinfo{person}{Christine Bauer}.} \bibinfo{year}{2023}\natexlab{}.
\newblock \showarticletitle{Amplifying Artists' Voices: Item Provider
  Perspectives on Influence and Fairness of Music Streaming Platforms}. In
  \bibinfo{booktitle}{\emph{Proceedings of the 31st {ACM} Conference on User
  Modeling, Adaptation and Personalization}} (Limassol, Cyprus)
  \emph{(\bibinfo{series}{UMAP '23})}. \bibinfo{publisher}{Association for
  Computing Machinery}, \bibinfo{address}{New York, NY, USA},
  \bibinfo{pages}{238--249}.
\newblock
\urldef\tempurl%
\url{https://doi.org/10.1145/3565472.3592960}
\showDOI{\tempurl}


\bibitem[Ekstrand(2020)]%
        {ekstrand2020_lenskit}
\bibfield{author}{\bibinfo{person}{Michael~D. Ekstrand}.}
  \bibinfo{year}{2020}\natexlab{}.
\newblock \showarticletitle{{LensKit} for {Python}: Next-Generation Software
  for Recommender Systems Experiments}. In
  \bibinfo{booktitle}{\emph{Proceedings of the 29th {ACM} International
  Conference on Information \& Knowledge Management}} (Virtual Event, Ireland)
  \emph{(\bibinfo{series}{CIKM '20})}. \bibinfo{publisher}{Association for
  Computing Machinery}, \bibinfo{address}{New York, NY, USA},
  \bibinfo{pages}{2999--3006}.
\newblock
\urldef\tempurl%
\url{https://doi.org/10.1145/3340531.3412778}
\showDOI{\tempurl}


\bibitem[Ekstrand et~al\mbox{.}(2024)]%
        {ekstrand2024distributions}
\bibfield{author}{\bibinfo{person}{Michael~D. Ekstrand}, \bibinfo{person}{Ben
  Carterette}, {and} \bibinfo{person}{Fernando Diaz}.}
  \bibinfo{year}{2024}\natexlab{}.
\newblock \showarticletitle{Distributionally-Informed Recommender System
  Evaluation}.
\newblock \bibinfo{journal}{\emph{{ACM} Transactions on Recommender Systems}}
  \bibinfo{volume}{2}, \bibinfo{number}{1}, Article \bibinfo{articleno}{6}
  (\bibinfo{date}{March} \bibinfo{year}{2024}), \bibinfo{numpages}{27}~pages.
\newblock
\urldef\tempurl%
\url{https://doi.org/10.1145/3613455}
\showDOI{\tempurl}


\bibitem[Ekstrand et~al\mbox{.}(2022a)]%
        {Ekstrand2022Fairness}
\bibfield{author}{\bibinfo{person}{Michael~D. Ekstrand},
  \bibinfo{person}{Anubrata Das}, \bibinfo{person}{Robin Burke}, {and}
  \bibinfo{person}{Fernando Diaz}.} \bibinfo{year}{2022}\natexlab{a}.
\newblock \showarticletitle{Fairness in Information Access Systems}.
\newblock \bibinfo{journal}{\emph{Foundations and
  Trends\textsuperscript{\textregistered} in Information Retrieval}}
  \bibinfo{volume}{16}, \bibinfo{number}{1--2} (\bibinfo{year}{2022}),
  \bibinfo{numpages}{196}~pages.
\newblock
\urldef\tempurl%
\url{https://doi.org/10.1561/1500000079}
\showDOI{\tempurl}


\bibitem[Ekstrand and Kluver(2021)]%
        {Ekstrand2021}
\bibfield{author}{\bibinfo{person}{Michael~D. Ekstrand} {and}
  \bibinfo{person}{Daniel Kluver}.} \bibinfo{year}{2021}\natexlab{}.
\newblock \showarticletitle{Exploring Author Gender in Book Rating and
  Recommendation}.
\newblock \bibinfo{journal}{\emph{User Modeling and User-Adapted Interaction}}
  \bibinfo{volume}{31}, \bibinfo{number}{3} (\bibinfo{year}{2021}),
  \bibinfo{pages}{377--420}.
\newblock
\urldef\tempurl%
\url{https://doi.org/10.1007/s11257-020-09284-2}
\showDOI{\tempurl}


\bibitem[Ekstrand and Mahant(2017)]%
        {ekstrandSturgeonCoolKids2017}
\bibfield{author}{\bibinfo{person}{Michael~D. Ekstrand} {and}
  \bibinfo{person}{Vaibhav Mahant}.} \bibinfo{year}{2017}\natexlab{}.
\newblock \showarticletitle{Sturgeon and the {{Cool Kids}}: {{Problems}} with
  {{Top-N Recommender Evaluation}}}. In \bibinfo{booktitle}{\emph{Proceedings
  of the 30th {{Florida Artificial Intelligence Research Society Conference}}}}
  (Marco Island, FL, USA) \emph{(\bibinfo{series}{FLAIRS 2017})}.
  \bibinfo{publisher}{AAAI Press}, \bibinfo{address}{Washington, DC, USA},
  \bibinfo{pages}{639--644}.
\newblock
\urldef\tempurl%
\url{https://aaai.org/papers/639-flairs-2017-15534/}
\showURL{%
\tempurl}


\bibitem[Ekstrand et~al\mbox{.}(2022b)]%
        {TREC2021}
\bibfield{author}{\bibinfo{person}{Michael~D Ekstrand}, \bibinfo{person}{Graham
  McDonald}, \bibinfo{person}{Amifa Raj}, {and} \bibinfo{person}{Isaac
  Johnson}.} \bibinfo{year}{2022}\natexlab{b}.
\newblock \showarticletitle{Overview of the {{TREC}} 2021 Fair Ranking Track}.
  In \bibinfo{booktitle}{\emph{The {{Thirtieth Text REtrieval Conference}}
  ({{TREC}} 2021) {{Proceedings}}}}.
\newblock
\urldef\tempurl%
\url{https://trec.nist.gov/pubs/trec30/papers/Overview-F.pdf}
\showURL{%
\tempurl}


\bibitem[Elahi et~al\mbox{.}(2021)]%
        {elahi2021investigating}
\bibfield{author}{\bibinfo{person}{Mehdi Elahi}, \bibinfo{person}{Danial~Khosh
  Kholgh}, \bibinfo{person}{Mohammad~Sina Kiarostami}, \bibinfo{person}{Sorush
  Saghari}, \bibinfo{person}{Shiva~Parsa Rad}, {and} \bibinfo{person}{Marko
  Tkal{\v{c}}i{\v{c}}}.} \bibinfo{year}{2021}\natexlab{}.
\newblock \showarticletitle{Investigating the Impact of Recommender Systems on
  User-based and Item-based Popularity Bias}.
\newblock \bibinfo{journal}{\emph{Information Processing \& Management}}
  \bibinfo{volume}{58}, \bibinfo{number}{5}, Article
  \bibinfo{articleno}{102655} (\bibinfo{date}{Sept.} \bibinfo{year}{2021}),
  \bibinfo{numpages}{15}~pages.
\newblock
\urldef\tempurl%
\url{https://doi.org/10.1016/j.ipm.2021.102655}
\showDOI{\tempurl}


\bibitem[Epps-Darling et~al\mbox{.}(2020)]%
        {EppsDarling2020Gender}
\bibfield{author}{\bibinfo{person}{Avriel Epps-Darling},
  \bibinfo{person}{Romain Takeo~Bouyer}, {and} \bibinfo{person}{Henriette
  Cramer}.} \bibinfo{year}{2020}\natexlab{}.
\newblock \showarticletitle{Artist Gender Representation in Music Streaming}.
  In \bibinfo{booktitle}{\emph{Proceedings of the 21st International Society
  for Music Information Retrieval Conference}} (Montréal, Canada)
  \emph{(\bibinfo{series}{ISMIR 2020})}. \bibinfo{publisher}{ISMIR},
  \bibinfo{pages}{248--254}.
\newblock
\urldef\tempurl%
\url{https://doi.org/10.5281/zenodo.4245416}
\showDOI{\tempurl}


\bibitem[Fabbri et~al\mbox{.}(2022)]%
        {Fabbri_Croci_Bonchi_Castillo_2022}
\bibfield{author}{\bibinfo{person}{Francesco Fabbri},
  \bibinfo{person}{Maria~Luisa Croci}, \bibinfo{person}{Francesco Bonchi},
  {and} \bibinfo{person}{Carlos Castillo}.} \bibinfo{year}{2022}\natexlab{}.
\newblock \showarticletitle{Exposure Inequality in People Recommender Systems:
  The Long-Term Effects}.
\newblock \bibinfo{journal}{\emph{Proceedings of the International {AAAI}
  Conference on Web and Social Media}} \bibinfo{volume}{16},
  \bibinfo{number}{1} (\bibinfo{date}{May} \bibinfo{year}{2022}),
  \bibinfo{pages}{194--204}.
\newblock
\urldef\tempurl%
\url{https://doi.org/10.1609/icwsm.v16i1.19284}
\showDOI{\tempurl}


\bibitem[Ferraro et~al\mbox{.}(2019)]%
        {ferraro2019}
\bibfield{author}{\bibinfo{person}{Andres Ferraro}, \bibinfo{person}{Dmitry
  Bogdanov}, \bibinfo{person}{Xavier Serra}, {and} \bibinfo{person}{Jason
  Yoon}.} \bibinfo{year}{2019}\natexlab{}.
\newblock \showarticletitle{Artist and Style Exposure Bias in Collaborative
  Filtering Based Music Recommendations}. In
  \bibinfo{booktitle}{\emph{Proceedings of the 1st Workshop on Designing
  Human-Centric Music Information Research Systems}} (Delft, The Netherlands)
  \emph{(\bibinfo{series}{wsHCMIR '19})}. \bibinfo{pages}{8--10}.
\newblock
\urldef\tempurl%
\url{https://doi.org/10.48550/arXiv.1911.04827}
\showDOI{\tempurl}


\bibitem[Ferraro et~al\mbox{.}(2021a)]%
        {ferraro2021_break}
\bibfield{author}{\bibinfo{person}{Andres Ferraro}, \bibinfo{person}{Xavier
  Serra}, {and} \bibinfo{person}{Christine Bauer}.}
  \bibinfo{year}{2021}\natexlab{a}.
\newblock \showarticletitle{Break the Loop: Gender Imbalance in Music
  Recommenders}. In \bibinfo{booktitle}{\emph{Proceedings of the 2021
  Conference on Human Information Interaction and Retrieval}} (Canberra, ACT,
  Australia) \emph{(\bibinfo{series}{CHIIR '21})}.
  \bibinfo{publisher}{Association for Computing Machinery},
  \bibinfo{address}{New York, NY, USA}, \bibinfo{pages}{249--254}.
\newblock
\urldef\tempurl%
\url{https://doi.org/10.1145/3406522.3446033}
\showDOI{\tempurl}


\bibitem[Ferraro et~al\mbox{.}(2021b)]%
        {Ferraro2021Fair}
\bibfield{author}{\bibinfo{person}{Andres Ferraro}, \bibinfo{person}{Xavier
  Serra}, {and} \bibinfo{person}{Christine Bauer}.}
  \bibinfo{year}{2021}\natexlab{b}.
\newblock \showarticletitle{What Is Fair? {E}xploring the Artists' Perspective
  on the Fairness of Music Streaming Platforms}. In
  \bibinfo{booktitle}{\emph{Human-Computer Interaction -- INTERACT 2021: 18th
  IFIP TC 13 International Conference}} (Bari, Italy)
  \emph{(\bibinfo{series}{INTERACT '21}, Vol.~\bibinfo{volume}{12933})}.
  \bibinfo{publisher}{Springer}, \bibinfo{address}{Cham, Switzerland},
  \bibinfo{pages}{562--584}.
\newblock
\urldef\tempurl%
\url{https://doi.org/10.1007/978-3-030-85616-8_33}
\showDOI{\tempurl}


\bibitem[Hazrati and Ricci(2022)]%
        {HAZRATI2022102766}
\bibfield{author}{\bibinfo{person}{Naieme Hazrati} {and}
  \bibinfo{person}{Francesco Ricci}.} \bibinfo{year}{2022}\natexlab{}.
\newblock \showarticletitle{Recommender Systems Effect on the Evolution of
  Users' Choices Distribution}.
\newblock \bibinfo{journal}{\emph{Information Processing \& Management}}
  \bibinfo{volume}{59}, \bibinfo{number}{1}, Article
  \bibinfo{articleno}{102766} (\bibinfo{year}{2022}),
  \bibinfo{numpages}{18}~pages.
\newblock
\urldef\tempurl%
\url{https://doi.org/10.1016/j.ipm.2021.102766}
\showDOI{\tempurl}


\bibitem[Hazrati and Ricci(2024)]%
        {Hazrati2024_choicemodels}
\bibfield{author}{\bibinfo{person}{Naieme Hazrati} {and}
  \bibinfo{person}{Francesco Ricci}.} \bibinfo{year}{2024}\natexlab{}.
\newblock \showarticletitle{Choice Models and Recommender Systems Effects on
  Users' Choices}.
\newblock \bibinfo{journal}{\emph{User Modeling and User-Adapted Interaction}}
  \bibinfo{volume}{34}, \bibinfo{number}{1} (\bibinfo{year}{2024}),
  \bibinfo{pages}{109--145}.
\newblock
\urldef\tempurl%
\url{https://doi.org/10.1007/s11257-023-09366-x}
\showDOI{\tempurl}


\bibitem[Hernandez et~al\mbox{.}(2022)]%
        {smith2022_inclusion_recording}
\bibfield{author}{\bibinfo{person}{Karla Hernandez}, \bibinfo{person}{Stacy~L.
  Smith}, \bibinfo{person}{Marc Choueiti}, {and} \bibinfo{person}{Katherine
  Pieper}.} \bibinfo{year}{2022}\natexlab{}.
\newblock \bibinfo{booktitle}{\emph{Inclusion in the Recording Studio?: Gender
  and Race/Ethnicity of Artists, Songwriters \& Producers across 1,000 Popular
  Songs from 2012--2021}}.
\newblock \bibinfo{type}{{T}echnical {R}eport}. \bibinfo{institution}{Annenberg
  Inclusion Initiative}.
\newblock
\urldef\tempurl%
\url{https://assets.uscannenberg.org/docs/aii-inclusion-recording-studio-20220331.pdf}
\showURL{%
\tempurl}


\bibitem[Hu et~al\mbox{.}(2008)]%
        {Hu2008_ials}
\bibfield{author}{\bibinfo{person}{Yifan Hu}, \bibinfo{person}{Yehuda Koren},
  {and} \bibinfo{person}{Chris Volinsky}.} \bibinfo{year}{2008}\natexlab{}.
\newblock \showarticletitle{Collaborative Filtering for Implicit Feedback
  Datasets}. In \bibinfo{booktitle}{\emph{2008 Eighth {IEEE} International
  Conference on Data Mining}} (Pisa, Italy) \emph{(\bibinfo{series}{ICDM
  2008})}. \bibinfo{publisher}{IEEE}, \bibinfo{address}{New York, NY, USA},
  \bibinfo{pages}{263--272}.
\newblock
\urldef\tempurl%
\url{https://doi.org/10.1109/ICDM.2008.22}
\showDOI{\tempurl}


\bibitem[{IFPI}(2024)]%
        {IFPI2024}
\bibfield{author}{\bibinfo{person}{{IFPI}}.} \bibinfo{year}{2024}\natexlab{}.
\newblock \bibinfo{booktitle}{\emph{{{IFPI Global Music Report 2024}}: State of
  the Industry}}.
\newblock \bibinfo{type}{{T}echnical {R}eport}. \bibinfo{institution}{IFPI},
  \bibinfo{address}{London, UK}.
\newblock
\urldef\tempurl%
\url{https://ifpi-website-cms.s3.eu-west-2.amazonaws.com/IFPI_GMR_2024_State_of_the_Industry_db92a1c9c1.pdf}
\showURL{%
\tempurl}


\bibitem[Jannach et~al\mbox{.}(2015)]%
        {jannach2015recommenders}
\bibfield{author}{\bibinfo{person}{Dietmar Jannach}, \bibinfo{person}{Lukas
  Lerche}, \bibinfo{person}{Iman Kamehkhosh}, {and} \bibinfo{person}{Michael
  Jugovac}.} \bibinfo{year}{2015}\natexlab{}.
\newblock \showarticletitle{What Recommenders Recommend: an Analysis of
  Recommendation Biases and Possible Countermeasures}.
\newblock \bibinfo{journal}{\emph{User Modeling and User-Adapted Interaction}}
  \bibinfo{volume}{25}, \bibinfo{number}{5} (\bibinfo{year}{2015}),
  \bibinfo{pages}{427--491}.
\newblock
\urldef\tempurl%
\url{https://doi.org/10.1007/s11257-015-9165-3}
\showDOI{\tempurl}


\bibitem[Kowald and Lacic(2022)]%
        {Kowald2022_Popularity}
\bibfield{author}{\bibinfo{person}{Dominik Kowald} {and}
  \bibinfo{person}{Emanuel Lacic}.} \bibinfo{year}{2022}\natexlab{}.
\newblock \showarticletitle{Popularity Bias in Collaborative Filtering-Based
  Multimedia Recommender Systems}. In \bibinfo{booktitle}{\emph{Advances in
  Bias and Fairness in Information Retrieval}} \emph{(\bibinfo{series}{BIAS
  2022}, Vol.~\bibinfo{volume}{1610})}. \bibinfo{publisher}{Springer},
  \bibinfo{address}{Cham, Switzerland}, \bibinfo{pages}{1--11}.
\newblock
\urldef\tempurl%
\url{https://doi.org/10.1007/978-3-031-09316-6_1}
\showDOI{\tempurl}


\bibitem[Mansoury et~al\mbox{.}(2020)]%
        {Mansoury2020Feedback}
\bibfield{author}{\bibinfo{person}{Masoud Mansoury}, \bibinfo{person}{Himan
  Abdollahpouri}, \bibinfo{person}{Mykola Pechenizkiy},
  \bibinfo{person}{Bamshad Mobasher}, {and} \bibinfo{person}{Robin Burke}.}
  \bibinfo{year}{2020}\natexlab{}.
\newblock \showarticletitle{Feedback Loop and Bias Amplification in Recommender
  Systems}. In \bibinfo{booktitle}{\emph{Proceedings of the 29th {ACM}
  International Conference on Information \& Knowledge Management}} (Virtual
  Event, Ireland) \emph{(\bibinfo{series}{CIKM '20})}.
  \bibinfo{publisher}{Association for Computing Machinery},
  \bibinfo{address}{New York, NY, USA}, \bibinfo{pages}{2145--2148}.
\newblock
\showISBNx{9781450368599}
\urldef\tempurl%
\url{https://doi.org/10.1145/3340531.3412152}
\showDOI{\tempurl}


\bibitem[Mitchell et~al\mbox{.}(2021)]%
        {mitchellAlgorithmicFairnessChoices2021}
\bibfield{author}{\bibinfo{person}{Shira Mitchell}, \bibinfo{person}{Eric
  Potash}, \bibinfo{person}{Solon Barocas}, \bibinfo{person}{Alexander
  D'Amour}, {and} \bibinfo{person}{Kristian Lum}.}
  \bibinfo{year}{2021}\natexlab{}.
\newblock \showarticletitle{Algorithmic Fairness: {Choices}, Assumptions, and
  Definitions}.
\newblock \bibinfo{journal}{\emph{Annual Review of Statistics and Its
  Application}}  \bibinfo{volume}{8} (\bibinfo{date}{Nov.}
  \bibinfo{year}{2021}), \bibinfo{pages}{141--163}.
\newblock
\urldef\tempurl%
\url{https://doi.org/10.1146/annurev-statistics-042720-125902}
\showDOI{\tempurl}


\bibitem[Raj and Ekstrand(2023)]%
        {rajUnifiedBrowsingModels2023}
\bibfield{author}{\bibinfo{person}{Amifa Raj} {and} \bibinfo{person}{Michael
  Ekstrand}.} \bibinfo{year}{2023}\natexlab{}.
\newblock \showarticletitle{Unified {{Browsing Models}} for {{Linear}} and
  {{Grid Layouts}}}.
\newblock \bibinfo{journal}{\emph{CoRR}} \bibinfo{number}{arXiv:2310.12524}
  (\bibinfo{year}{2023}), \bibinfo{numpages}{7}~pages.
\newblock
\urldef\tempurl%
\url{https://doi.org/10.48550/arXiv.2310.12524}
\showDOI{\tempurl}


\bibitem[Raj and Ekstrand(2022)]%
        {raj2022:ranking}
\bibfield{author}{\bibinfo{person}{Amifa Raj} {and} \bibinfo{person}{Michael~D
  Ekstrand}.} \bibinfo{year}{2022}\natexlab{}.
\newblock \showarticletitle{Measuring Fairness in Ranked Results: An Analytical
  and Empirical Comparison}. In \bibinfo{booktitle}{\emph{Proceedings of the
  45th International {ACM} {SIGIR} Conference on Research and Development in
  Information Retrieval}} (Madrid, Spain) \emph{(\bibinfo{series}{SIGIR '22})}.
  \bibinfo{publisher}{Association for Computing Machinery},
  \bibinfo{address}{New York, NY, USA}, \bibinfo{pages}{726--736}.
\newblock
\urldef\tempurl%
\url{https://doi.org/10.1145/3477495.3532018}
\showDOI{\tempurl}


\bibitem[Raj and Ekstrand(2024)]%
        {rajOptimizingRankingGridlayout2024}
\bibfield{author}{\bibinfo{person}{Amifa Raj} {and} \bibinfo{person}{Michael~D.
  Ekstrand}.} \bibinfo{year}{2024}\natexlab{}.
\newblock \showarticletitle{Towards Optimizing Ranking in Grid-Layout for
  Provider-Side Fairness}. In \bibinfo{booktitle}{\emph{Advances in
  {{Information Retrieval}}}} (Glasgow, UK) \emph{(\bibinfo{series}{ECIR 2024},
  Vol.~\bibinfo{volume}{14612})}. \bibinfo{publisher}{Springer},
  \bibinfo{address}{Cham, Switzerland}, \bibinfo{pages}{90--105}.
\newblock
\urldef\tempurl%
\url{https://doi.org/10.1007/978-3-031-56069-9_7}
\showDOI{\tempurl}


\bibitem[Rendle et~al\mbox{.}(2009)]%
        {rendleBPRBayesianPersonalized2009}
\bibfield{author}{\bibinfo{person}{Steffen Rendle}, \bibinfo{person}{Christoph
  Freudenthaler}, \bibinfo{person}{Zeno Gantner}, {and} \bibinfo{person}{Lars
  {Schmidt-Thieme}}.} \bibinfo{year}{2009}\natexlab{}.
\newblock \showarticletitle{{{BPR}}: {{Bayesian Personalized Ranking}} from
  Implicit Feedback}. In \bibinfo{booktitle}{\emph{Proceedings of the
  Twenty-Fifth Conference on Uncertainty in Artificial Intelligence}}
  (Montreal, Quebec, Canada) \emph{(\bibinfo{series}{{{UAI}} '09})}.
  \bibinfo{publisher}{AUAI Press}, \bibinfo{address}{Arlington, VA, USA},
  \bibinfo{pages}{452--461}.
\newblock
\urldef\tempurl%
\url{https://doi.org/10.5555/1795114.1795167}
\showDOI{\tempurl}


\bibitem[Sapiezynski et~al\mbox{.}(2019)]%
        {AWRF}
\bibfield{author}{\bibinfo{person}{Piotr Sapiezynski}, \bibinfo{person}{Wesley
  Zeng}, \bibinfo{person}{Ronald E~Robertson}, \bibinfo{person}{Alan Mislove},
  {and} \bibinfo{person}{Christo Wilson}.} \bibinfo{year}{2019}\natexlab{}.
\newblock \showarticletitle{Quantifying the Impact of User Attentionon Fair
  Group Representation in Ranked Lists}. In \bibinfo{booktitle}{\emph{Companion
  Proceedings of The 2019 World Wide Web Conference}} (San Francisco, CA, USA)
  \emph{(\bibinfo{series}{{{WWW}} '19 {{Companion}}})}.
  \bibinfo{publisher}{Association for Computing Machinery},
  \bibinfo{address}{New York, NY, USA}, \bibinfo{pages}{553--562}.
\newblock
\urldef\tempurl%
\url{https://doi.org/10.1145/3308560.3317595}
\showDOI{\tempurl}


\bibitem[Schedl et~al\mbox{.}(2022a)]%
        {schedl2022_lfm2b}
\bibfield{author}{\bibinfo{person}{Markus Schedl}, \bibinfo{person}{Stefan
  Brandl}, \bibinfo{person}{Oleg Lesota}, \bibinfo{person}{Emilia
  Parada-Cabaleiro}, \bibinfo{person}{David Penz}, {and} \bibinfo{person}{Navid
  Rekabsaz}.} \bibinfo{year}{2022}\natexlab{a}.
\newblock \showarticletitle{{LFM-2b}: A Dataset of Enriched Music Listening
  Events for Recommender Systems Research and Fairness Analysis}. In
  \bibinfo{booktitle}{\emph{Proceedings of the 2022 Conference on Human
  Information Interaction and Retrieval}} (Regensburg, Germany)
  \emph{(\bibinfo{series}{CHIIR '22})}. \bibinfo{publisher}{Association for
  Computing Machinery}, \bibinfo{address}{New York, NY, USA},
  \bibinfo{pages}{337--341}.
\newblock
\showISBNx{9781450391863}
\urldef\tempurl%
\url{https://doi.org/10.1145/3498366.3505791}
\showDOI{\tempurl}


\bibitem[Schedl et~al\mbox{.}(2022b)]%
        {schedl2022_handbook}
\bibfield{author}{\bibinfo{person}{Markus Schedl}, \bibinfo{person}{Peter
  Knees}, \bibinfo{person}{Brian McFee}, {and} \bibinfo{person}{Dmitry
  Bogdanov}.} \bibinfo{year}{2022}\natexlab{b}.
\newblock \showarticletitle{Music Recommendation Systems: {Techniques}, Use
  Cases, and Challenges}.
\newblock In \bibinfo{booktitle}{\emph{Recommender Systems Handbook}
  (\bibinfo{edition}{3rd} ed.)}, \bibfield{editor}{\bibinfo{person}{Francesco
  Ricci}, \bibinfo{person}{Lior Rokach}, {and} \bibinfo{person}{Bracha
  Shapira}} (Eds.). \bibinfo{publisher}{Springer US}, \bibinfo{address}{New
  York, NY, USA}, \bibinfo{pages}{927--971}.
\newblock
\showISBNx{978-1-0716-2197-4}
\urldef\tempurl%
\url{https://doi.org/10.1007/978-1-0716-2197-4_24}
\showDOI{\tempurl}


\bibitem[Smith et~al\mbox{.}(2024)]%
        {smith2024_inclusion_recording}
\bibfield{author}{\bibinfo{person}{Stacy~L. Smith}, \bibinfo{person}{Katherine
  Pieper}, \bibinfo{person}{Karla Hernandez}, {and} \bibinfo{person}{Sam
  Wheeler}.} \bibinfo{year}{2024}\natexlab{}.
\newblock \bibinfo{booktitle}{\emph{Inclusion in the Recording Studio?: Gender
  and Race/Ethnicity of Artists, Songwriters \& Producers across 1,200 Popular
  Songs from 2012--2023}}.
\newblock \bibinfo{type}{{T}echnical {R}eport}. \bibinfo{institution}{Annenberg
  Inclusion Initiative}.
\newblock
\urldef\tempurl%
\url{https://assets.uscannenberg.org/docs/aii-inclusion-recording-studio-20240130.pdf}
\showURL{%
\tempurl}


\bibitem[Sonboli et~al\mbox{.}(2020a)]%
        {sonboliOpportunisticMultiaspectFairness2020a}
\bibfield{author}{\bibinfo{person}{Nasim Sonboli}, \bibinfo{person}{Farzad
  Eskandanian}, \bibinfo{person}{Robin Burke}, \bibinfo{person}{Weiwen Liu},
  {and} \bibinfo{person}{Bamshad Mobasher}.} \bibinfo{year}{2020}\natexlab{a}.
\newblock \showarticletitle{Opportunistic {{Multi-aspect Fairness}} through
  {{Personalized Re-ranking}}}. In \bibinfo{booktitle}{\emph{{{UMAP}} '20}}.
  \bibinfo{publisher}{Association for Computing Machinery},
  \bibinfo{address}{New York, NY, USA}, \bibinfo{pages}{239--247}.
\newblock
\urldef\tempurl%
\url{https://doi.org/10.1145/3340631.3394846}
\showDOI{\tempurl}


\bibitem[Sonboli et~al\mbox{.}(2020b)]%
        {Sonboli2020opportunistic}
\bibfield{author}{\bibinfo{person}{Nasim Sonboli}, \bibinfo{person}{Farzad
  Eskandanian}, \bibinfo{person}{Robin Burke}, \bibinfo{person}{Weiwen Liu},
  {and} \bibinfo{person}{Bamshad Mobasher}.} \bibinfo{year}{2020}\natexlab{b}.
\newblock \showarticletitle{Opportunistic Multi-aspect Fairness through
  Personalized Re-ranking}. In \bibinfo{booktitle}{\emph{Proceedings of the
  28th {ACM} Conference on User Modeling, Adaptation and Personalization}}
  (Genoa, Italy) \emph{(\bibinfo{series}{UMAP '20})}.
  \bibinfo{publisher}{Association for Computing Machinery},
  \bibinfo{address}{New York, NY, USA}, \bibinfo{pages}{239--247}.
\newblock
\urldef\tempurl%
\url{https://doi.org/10.1145/3340631.3394846}
\showDOI{\tempurl}


\bibitem[Wang and Horv{\'a}t(2019)]%
        {wang2019gender}
\bibfield{author}{\bibinfo{person}{Yixue Wang} {and}
  \bibinfo{person}{Em{\H{o}}ke-{\'A}gnes Horv{\'a}t}.}
  \bibinfo{year}{2019}\natexlab{}.
\newblock \showarticletitle{Gender Differences in the Global Music Industry:
  Evidence from {M}usicBrainz and {T}he {E}cho {N}est}.
\newblock \bibinfo{journal}{\emph{Proceedings of the International {AAAI}
  Conference on Web and Social Media}} \bibinfo{volume}{13},
  \bibinfo{number}{01} (\bibinfo{date}{July} \bibinfo{year}{2019}),
  \bibinfo{pages}{517--526}.
\newblock
\urldef\tempurl%
\url{https://doi.org/10.1609/icwsm.v13i01.3249}
\showDOI{\tempurl}


\bibitem[Wang et~al\mbox{.}(2023)]%
        {Wang2023_fairness}
\bibfield{author}{\bibinfo{person}{Yifan Wang}, \bibinfo{person}{Weizhi Ma},
  \bibinfo{person}{Min Zhang}, \bibinfo{person}{Yiqun Liu}, {and}
  \bibinfo{person}{Shaoping Ma}.} \bibinfo{year}{2023}\natexlab{}.
\newblock \showarticletitle{A Survey on the Fairness of Recommender Systems}.
\newblock \bibinfo{journal}{\emph{{ACM} Transactions on Information Systems}}
  \bibinfo{volume}{41}, \bibinfo{number}{3}, Article \bibinfo{articleno}{52}
  (\bibinfo{date}{Feb.} \bibinfo{year}{2023}), \bibinfo{numpages}{43}~pages.
\newblock
\urldef\tempurl%
\url{https://doi.org/10.1145/3547333}
\showDOI{\tempurl}


\bibitem[Watson(2020)]%
        {watsonprogramming}
\bibfield{author}{\bibinfo{person}{Jada Watson}.}
  \bibinfo{year}{2020}\natexlab{}.
\newblock \showarticletitle{Programming Inequality: Gender Representation on
  Canadian Country Radio (2005--2019)}. In
  \bibinfo{booktitle}{\emph{Proceedings of the 21st International Society for
  Music Information Retrieval Conference}} (Montréal, Canada)
  \emph{(\bibinfo{series}{ISMIR 2020})}. \bibinfo{publisher}{ISMIR},
  \bibinfo{pages}{392--399}.
\newblock
\urldef\tempurl%
\url{https://doi.org/10.5281/zenodo.4245452}
\showDOI{\tempurl}


\bibitem[Zehlike et~al\mbox{.}(2017)]%
        {zehlikeFAIRFair2017}
\bibfield{author}{\bibinfo{person}{Meike Zehlike}, \bibinfo{person}{Francesco
  Bonchi}, \bibinfo{person}{Carlos Castillo}, \bibinfo{person}{Sara Hajian},
  \bibinfo{person}{Mohamed Megahed}, {and} \bibinfo{person}{Ricardo
  {Baeza-Yates}}.} \bibinfo{year}{2017}\natexlab{}.
\newblock \showarticletitle{{{FA}}*{{IR}}: {{A Fair Top-k Ranking Algorithm}}}.
  In \bibinfo{booktitle}{\emph{Proceedings of the 2017 {ACM} on Conference on
  Information and Knowledge Management}} (Singapore)
  \emph{(\bibinfo{series}{CIKM '17})}. \bibinfo{publisher}{Association for
  Computing Machinery}, \bibinfo{address}{New York, NY, USA},
  \bibinfo{pages}{1569--1578}.
\newblock
\urldef\tempurl%
\url{https://doi.org/10.1145/3132847.3132938}
\showDOI{\tempurl}


\bibitem[Zhang et~al\mbox{.}(2020)]%
        {zhang2020consumption}
\bibfield{author}{\bibinfo{person}{Jingjng Zhang}, \bibinfo{person}{Gediminas
  Adomavicius}, \bibinfo{person}{Alok Gupta}, {and} \bibinfo{person}{Wolfgang
  Ketter}.} \bibinfo{year}{2020}\natexlab{}.
\newblock \showarticletitle{Consumption and Performance: Understanding
  Longitudinal Dynamics of Recommender Systems via an Agent-Based Simulation
  Framework}.
\newblock \bibinfo{journal}{\emph{Information Systems Research}}
  \bibinfo{volume}{31}, \bibinfo{number}{1} (\bibinfo{year}{2020}),
  \bibinfo{pages}{76--101}.
\newblock
\urldef\tempurl%
\url{https://doi.org/10.1287/isre.2019.0876}
\showDOI{\tempurl}


\end{thebibliography}

\end{document}